# A framework for the comparison of different EEG acquisition solutions

Aurore Bussalb [(1)], PhD student, Marie Prat [(1)], David Ojeda [(1)], PhD, Quentin Barthélemy [(1)], PhD, Julien Bonnaud [(1)], and Louis Mayaud [(1)], D.Phil.

(1) Mensia Technologies, France.

The purpose of this work is to propose a framework for the benchmarking of EEG amplifiers, headsets, and electrodes providing objective recommendation for a given application. The framework covers: data collection paradigm, data analysis, and statistical framework.

To illustrate, data was collected from 12 different devices totaling up to 6 subjects per device. Two data acquisition protocols were implemented: a resting-state protocol eyes-open (EO) and eyes-closed (EC), and an Auditory Evoked Potential (AEP) protocol. Signal-to-noise ratio (SNR) on alpha band (EO/EC) and Event Related Potential (ERP) were extracted as objective quantification of physiologically meaningful information. Then, visual representation, univariate statistical analysis, and multivariate model were performed to increase results interpretability.

Objective criteria show that the spectral SNR in alpha does not provide much discrimination between systems, suggesting that the acquisition quality might not be of primary importance for spectral and specifically alpha-based applications. On the contrary, AEP SNR proved much more variable stressing the importance of the acquisition setting for ERP experiments. The multivariate analysis identified some individuals and some systems as independent statistically significant contributors to the SNR. It highlights the importance of inter-individual differences in neurophysiological experiments (sample size) and suggests some device might objectively be superior to others when it comes to ERP recordings.

However, the illustration of the proposed benchmarking framework suffers from severe limitations including small sample size and sound card jitter in the auditory stimulations. While these limitations hinders a definite ranking of the evaluated hardware, we believe the proposed benchmarking framework to be a modest yet valuable contribution to the field.

# Table of Contents

# Introduction

The existence of electrical brain currents in the brain was discovered in 1875 by Richard Caton (Haas, 2003). Fifty years later, Hans Berger, a German neurologist, recorded brain electrical activity measured on the human scalp. He noticed that the activity recorded is different according to the functional status of the brain (sleep, epilepsy, ...) (Berger, 1929). Since then, research on neurophysiology greatly expanded and electroencephalograms (EEG) have somewhat incrementally improved, eventually turning it into a modern neuroimaging technique (Teplan, 2002). EEG is recorded on the scalp with electrodes that convert the body's bioelectricity into analogous signals then amplified and digitized. The targeted bioelectricity in EEG is generated by cortical displacement of charged particles reflecting instantaneous brain function. It is often contaminated by peripheral muscle activity (EMG) from the face and neck. In addition to that, Electromagnetic (EM) perturbations are typically lowering further the signal-to-noise ratio (SNR). Robustness to external sources of contaminations is an essential part of an acquisition system that influences: electrode material, analogue signal propagation, shielding, signal amplification, and digitalization.

The recent years have seen a boom in existing EEG recordings systems. While mostly similar from the technological standpoint, they can differ on (a) the number, place, and type (dry or wet) of electrodes used, (b) the protection against power lines interferences, and (c) amplification and digitalization strategies. The choice of the system for an experiment typically depends on the purpose of the application (general consumer, clinical or research) and where the recording takes place (hospital, research lab or the home), which is why the compliance with regulatory norms (ISO 60601-2-26 for instance) of a given system often matters. Beyond this, the performance is usually solely informed by technical specifications that are reported by the manufacturer (noise level, input impedance). These quality indicators do not necessarily relate to actual ecological performance. For instance a dry system might exhibit excellent technical specifications but turn out extremely sensitive to movement artefacts that will not be captured by technical specifications. In addition to this, there is quite large price variety between systems and it is not clear how it relates to performance. For these reasons, we suggest a framework to objectively assess the ecological performance of these systems with respect to physiologically meaningful criteria.

To illustrate this framework, data was collected using NeuroRT Harvest$^{TM}$ (Mensia Technologies, Paris, France), a software developed for an integrated data collection process across several physical sites. The resulting data was then analyzed regarding to objective data quality criteria so as to extract meaningful recommendations.

# Material and methods

## Acquisition Procedure

NeuroRT Harvest[TM] is a standalone application for the acquisition of physiological time series. It features a user-friendly API to design acquisition protocols as well as standalone application to guide layperson through the data collection protocol, record EEG signal, and store them remotely with consistent meta-information (including unique subject ID).

In this experiment, EEGs were recorded following two standard protocols implemented with NeuroRT Harvest[TM]:

- A resting state protocol made of three minutes recorded with eyes open (EO) and three minutes with eyes closed (EC);

- An Auditory Evoked Potentials (AEPs)[1] (Jeweet et al., 1970) protocol made of 3 minutes of AEPs recorded with eyes open. In this protocol, sinusoidal auditory beeps ("targets") at 1kHz were played for 100ms every 500ms.

Data was collected from 12 different devices featuring dry, saline or gel electrodes. Participants included 19 healthy volunteers located in two distant physical locations: 4 women and 15 men.

## Data Processing

Data was processed using offline methods implemented with NeuroRT Studio[TM] (v. 3.1, Mensia Technologies, Paris, France):

- EEG signals collected during the "Resting state EO-EC" protocol were first all downsampled at 128Hz. Then signals were filtered with a 3$^{rd}$-order Butterworth filter in the frequency range 0.1-40Hz. Eventually, spectrum was extracted using the Welch method, i.e. applying the Fast Fourier Transform (FFT) on a 2 second-long window epoched every 125ms after a Hann window had been applied, and then averaging along the recording. The averaged spectrum over subjects was then normalized using the newly suggested 1/f normalization using a channel-wise robust linear regression on the log-log spectrum (Barthelemy et al.).

- EEG signals recorded during the "Auditory Evoked Potentials" protocol were also downsampled at 128Hz. Then a 2$^{nd}$-order Butterworth filter was applied in the frequency range 0.5-40Hz, followed by a band-stop filter in the frequency range 48-

---

[1] Also called Steady State Auditory Evoked Potentials (SSAEP), Auditory Steady State Evoked Potentials (ASSEP) or Auditory Steady State Responses (ASSR).

52Hz with a 4$^{th}$-order Butterworth notch filter to suppress 50Hz powerline interference. Eventually, signals were normalized on a one-second-long window using a Frobenius norm (defined as the square root of the sum of the absolute squares of its elements) applied across all samples and electrodes. In parallel, fake stimulations drawn from a Poisson distribution (the mean of which was set to match auditory stimulations frequency) were used to get a reference ERP. This process resulted in a set of preprocessed signals with real stimulations and one with fake stimulations equal in numbers.

# Objective Criteria

## *Spectral SNR*

For spectral analysis, the difference in alpha bandwidth was computed between EC and EO conditions. Alpha waves can be best seen with eyes closed and under conditions of physical relation and mental inactivity (Niedermeyer, 2005). Thus, we expected to observe a higher peak in the alpha band during EC condition.

Spectrum had been averaged over all subjects and then normalized with 1/f procedure. The next step was to compute the ratio between the mean power during EC ($\alpha_{eyes\ closed}$) and the mean power during EO ($\alpha_{eyes\ open}$) in the broadband alpha power defined between 6 and 14 Hz:

$$\text{SNR}_{spectral} = \frac{\alpha_{eyes\ closed}}{\alpha_{eyes\ open}}.$$

We obtained a ratio for each electrode available, then averaged to obtain one spectral SNR per headset. A spectral SNR higher than 1 was expected.

## *Auditory evoked potentials*

Several ERPs can be identified following an auditory stimulation. We chose to focus on the P100 component occurring around 100ms. This component was chosen over the P300 because of its low inter-individual variability as any exogenous ERP (Shagass et al., 1972) and because it can be analysed without unusually high sampling frequency, which lower latency ERP would require.

AEPs were plotted on one-second-long window (from zero, corresponding to the stimulation time, to one second). The SNR was extracted on a 100ms (between 50ms and 150ms) window that was corrected manually for the sound card offset (visually determined for each headset). This was possible thanks to constant polarity and relative temporal consistency of the P100 without affecting our results that does not take into account the exact latency of the P100.

First, the power of real and fake AEP were computed for each subject at each electrode available. Then, the ratio between the power in real AEP (AEP $_{real\ stimulation}$) and in fake AEP (AEP $_{fake\ stimulation}$) was computed for each subject at each electrode:

$$\text{SNR}_{AEP} = \frac{AEP_{real\ stimulation}}{AEP_{fake\ stimulations}}.$$

Eventually the average across all subjects and electrodes was computed to get one AEP SNR per headset. Since it was shown that auditory P100 component is more visible over frontal electrode sites (Brett-Green et al., 2008), AEP SNR per headset were also computed by averaging only results across frontal electrodes.

## *Univariate analysis*

Statistical analysis were run using Python Scipy Library (version 0.12.0). The following statistical tests were based on the spectral and AEP SNR computed per subject and averaged across all electrodes.

A Shapiro-Wilk test (Shapiro et al., 1965) was run to check for normality to allow for the use of parametric tests. Then, the Levene test (Levene, 1960) was performed to test for homoscedasticity.

If these two assumptions were satisfied, the amplifier effect was assessed thanks to a one-way analysis of variance (ANOVA) (Heiman, 2002) where the system was the independent variable. The null hypothesis tested here was that for a given SNR variable, mean values in every amplifier were the same:

$$H_0: \mu_A = \mu_B = ... = \mu_L,$$

where $\mu_X$ corresponds to the mean SNR across subjects in system X.

When distributions of the quantitative variables could not be approximated by normal distributions, non-parametric tests were conducted using the Kruskal-Wallis (KW) tests (Kruskal et al., 1952) to assess the effect of amplifiers based on the same null hypothesis as for the ANOVA.

## *Multivariate analysis*

In several cases, one subject had their EEG recorded for different headsets and headsets where investigated with potentially different subsets of subjects, hereby creating a potential bias. To discard this bias we attempted to correct for potential "physiological outliers", i.e. subject with consistently low or high SNR across devices. Using a multivariate analysis where independent variables dummy-coded for device and subject to explain SNR could potentially correct for this bias while identifying statistically significant independent contribution of specific devices.

Multivariate analysis was performed using a Least Absolute Shrinkage and Selection Operator (LASSO) (Tibshirani, 1996) implemented in Python using the Akaike Information Criterion (AIC) for the selection of optimal value for the regularization parameter (lambda). LASSO minimizes the usual sum of squared errors, with a penalty on the absolute values of the coefficients (l1-norm) and performs variable selection as it allows to set coefficients to zero exactly. This way, it favors the emergence of predictive variables and improve the model interpretation:

$$\hat{\beta} = argmin_\beta \sum_{i=1}^{N}(y_i - \hat{y}_i)^2 + \lambda\,|\beta|,$$

where $y_i$ corresponds to the dependent variables (AEP or spectral SNR), $y_i$ *hat* the predicted values, *lambda* the tuning parameter, *bêta* the coefficients found by linear regression associated to the independent variables (the headsets and subjects). Two multivariate analysis were run: first where the spectral SNR was the dependent variable and second where the AEP SNR was. In both cases, the independent variables were the subjects and the headsets coded as dummies.

# Results

## Spectral analysis

Spectral SNRs per headset were computed from eyes open and eyes closed spectra after normalization as shown in Figure 1 and Table 1.

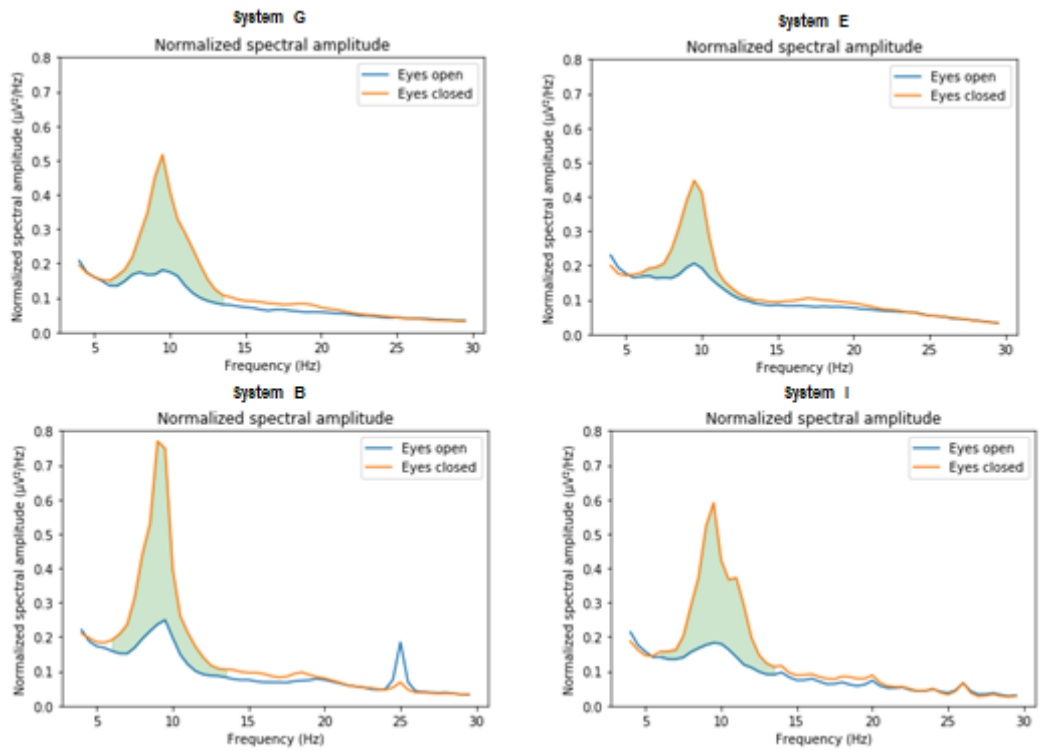

**Figure 1.** Mean spectrum over subjects and electrodes for four EEG systems. The green area represents the spectral SNR.

**Table 1.** Mean spectral SNR values extracted from spectral analysis between eyes-open (EO) and eyes-closed (EC) conditions. System names indicated in bold are selected as independent contributors to SNR after correcting for users recorded.

| System | Number of subjects recorded | Mean of spectral SNR over all electrodes (±std) |
| --- | --- | --- |
| System I | 5 | 2.00±0.30 |
| System B | 3 | 1.99±0.49 |
| System A | 4 | 1.96±0.68 |
| System H | 5 | 1.94±0.26 |
| System J | 2 | 1.92±0.32 |
| **System K** | 5 | 1.89±0.59 |
| System D | 4 | 1.88±0.28 |
| System G | 6 | 1.87±0.16 |
| System C | 5 | 1.75±0..20 |

| System F | 5 | 1.61±0.20 |
| System E | 5 | 1.49±0.11 |
| System L | 5 | 1.46±0.21 |

## Auditory evoked potential analysis

As expected, AEP plots show a positive peak at approximately 100ms after the auditory stimulus at 0s as illustrated in Figure 2. This result is consistent with Čeponien et al., 1998 and Brett-Green et al., 2008 also showing a P100 on frontal electrodes after a frequent auditory stimulation.

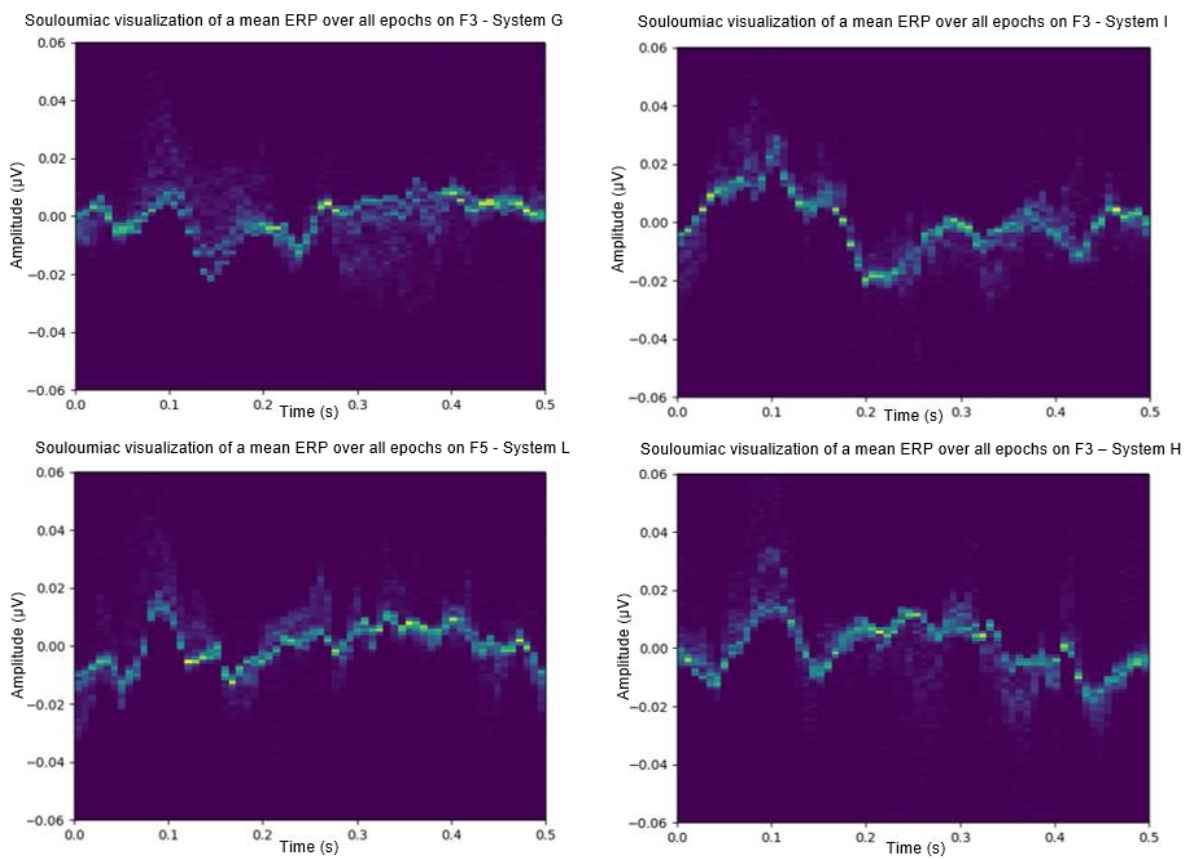

**Figure 2.** P100 ERPs density map (Souloumiac et Rivet 2013) for all epochs from all subjects.

AEP SNR are presented in Table 2.

**Table 2.** Mean SNR values of AEP over all and frontal electrodes computed according to the sound card offset. System names indicated in bold are selected as independent contributors to SNR after correcting for users recorded.

| System Name | Number of subject recorded | Mean of AEP SNR over all electrodes (±std) | Mean of AEP SNR over frontal electrodes (±std) |
|---|---|---|---|
| **System L** | 5 | 3.23±2.19 | 3.13±1.48 |
| **System G** | 6 | 2.40±1.46 | 3.51±2.35 |
| System E | 5 | 1.79±0.56 | 1.73±0.61 |
| System I | 4 | 1.62±0.85 | 2.01±0.86 |
| System K | 5 | 1.62±0.95 | 1.93±1.21 |
| System J | 2 | 1.56±0.91 | 1.09±0.55 |
| System A | 4 | 1.32±0.89 | 1.00±0.58 |
| System D | 4 | 1.30±1.06 | 0.91±0.79 |
| System B | 3 | 1.26±0.48 | 1.77±0.16 |
| System H | 5 | 1.21±0.45 | 1.25±0.54 |
| System C | 5 | 1.16±0.44 | 1.06±0.30 |
| **System F** | 5 | 0.62±0.33 | 1.09±0.070 |

## Univariate analysis

Levene test confirmed the equality of variances but according to Shapiro-Wilk normality test, neither spectral SNR nor AEP SNR followed a normal distribution. So KW tests were run and they failed to reject the null hypothesis that the mean SNR across subjects were not significantly different between systems in spectral SNR (p-value = 0.99) and AEP SNR (p-value = 0.19).

## Multivariate analysis

First, the Lasso was performed on spectral SNR values as dependent variables. With a lambda found equal to 0.020 with AIC criterion, two variables with non-zero coefficient were returned by the model:

- one subject, associated with good spectral SNR, indicating that it showed a consistently higher SNR across different system recording;
- the system K that was also associated with good spectral SNR consistently across different users.

Regarding the AEP SNR, the lambda found by AIC criterion was equal to 0.016 and led to the selection of seven variables:

- four subjects among which three were associated with good AEP SNR and one with bad;
- three systems: G and L associated with good AEP SNR at the opposite of F; these results are consistent with those presented in Table 2.

# Discussion

The data collection was designed to acquire a consistent dataset with different EEG systems using NeuroRT Harvest[TM], in order to further make devices comparison based on objective criteria that were: signal-to-noise ratio between eyes open and eyes closed conditions and AEP shapes following auditory stimuli.

In order to increase the precision of the spectral SNR, we applied the 1/f normalization (with the exclusion of the alpha band) to the mean spectrum as illustrated in Figure 5.

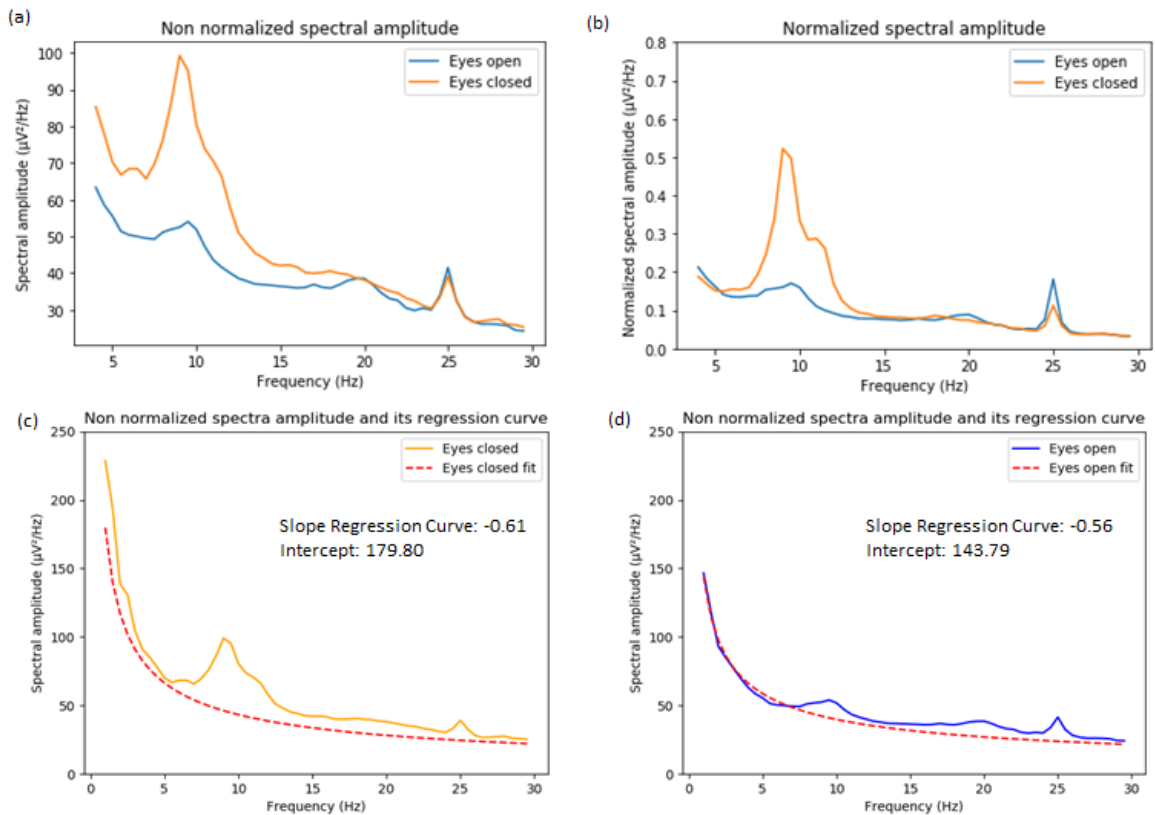

**Figure 3.** Comparison between mean spectra across subjects and electrodes for System J before (a) and after (b) 1/f normalization. The regression curve and the coefficients are shown for eyes closed (c) and eyes open (d) conditions.

Multivariate analysis revealed that little significant differences were observed between headsets, especially regarding the spectral SNR, results confirmed by the univariate analysis. This was partly due to small population size and the high intra- and inter-individual variability. The LASSO confirmed that several subjects explained a significant part of the variance making these results difficult to interpret and suggesting a more extensive data collection effort is required.

Some devices appeared to provide sensibly better signal quality regarding spectral SNR and AEP SNR even if the differences were not significant.

Interpretation of these results must be considered in the light of the study limitations. Especially, sound card latency should be investigated in order to reduce the delay between the stimulation and the actual auditory beep to less than 50ms. Finally, data should be acquired on a larger population size in order to give more power to statistical tests applied on data.

# Conclusion

We proposed here a framework for the objective comparison of different EEG systems. This framework covered the methodological protocol (number of subjects, number of recordings, type of recording), the acquisition protocol (length of recording, stimulation type), the data processing methods (pre-processing, denoising, and signal quality indices), and the statistical framework to discard external sources of variance and concentrate on the influence of each device. We illustrated this framework with the comparison of 12 different headsets, which suffered significant limitations. Primarily, the number of subjects recorded was too small to lead to significant conclusion. We believe however that the methods and framework described here remain of interest to the community.

# Acknowledgments

We would like to precise that this work was conducted partly thanks the EU H2020 NEWROFEED grant 684809, MindYourBrain and Reminary projects.

# References

Haas, L. F. (2003). Hans Berger (1873–1941), Richard Caton (1842–1926), and electroencephalography. *Journal of Neurology, Neurosurgery & Psychiatry*, 74(1), 9-9.

Berger, H. (1929). Über das elektrenkephalogramm des menschen. *Archiv für psychiatrie und nervenkrankheiten*, *87*(1), 527-570.

Teplan, M. (2002). Fundamentals of EEG measurement. *Measurement science review*, *2*(2), 1-11.

Jewett, D. L., Romano, M. N., & Williston, J. S. (1970). Human auditory evoked potentials: possible brain stem components detected on the scalp. *Science*, *167*(3924), 1517-1518.

Barthelemy, Q., Bussalb, A., Ojeda, D., Mayaud, L. (2019). Normalization of EEG spectra with 1/f regression. Manuscript submitted for publication.

Niedermeyer, E. (2005). The normal EEG of the waking adult. *Electroencephalography: Basic principles, clinical applications, and related fields*, *167*, 155-164.

Shagass, C. (1972). Characteristics of Event-Related Potentials. In *Evoked Brain Potentials in Psychiatry* (pp. 49-86). Springer, Boston, MA.

Brett-Green, B. A., Miller, L. J., Gavin, W. J., & Davies, P. L. (2008). Multisensory integration in children: a preliminary ERP study. *Brain Research*, *1242*, 283-290.

Shapiro, S. S. & Wilk, M.B (1965). An analysis of variance test for normality (complete samples). *Biometrika*, Vol. 52, pp. 591-611.

Levene, H. (1960). *In Contributions to Probability and Statistics: Essays in Honor of Harold Hotelling*, I. Olkin et al. eds., Stanford University Press, pp. 278-292.

Heiman, G.W. (2002) *Research Methods in Statistics*.

Kruskal, W. H., & Wallis, W. A. (1952). Use of ranks in one-criterion variance analysis. *Journal of the American statistical Association*, 47(260), 583-621.

Tibshirani, R. (1996). Regression shrinkage and selection via the lasso. *Journal of the Royal Statistical Society. Series B (Methodological)*, 267-288.

Čeponien, R., Cheour, M., & Näätänen, R. (1998). Interstimulus interval and auditory event-related potentials in children: evidence for multiple generators. *Electroencephalography and Clinical Neurophysiology/Evoked Potentials Section*, *108*(4), 345-354.

Rivet, B., & Souloumiac, A. (2013). Optimal linear spatial filters for event-related potentials based on a spatio-temporal model: Asymptotical performance analysis. *Signal Processing*, *93*(2), 387-398.